\documentclass[prl,showpacs,superscriptaddress,twocolumn,merge]{revtex4-1}

\usepackage{graphicx}
\usepackage{amsmath}
\usepackage{amssymb}
\usepackage{dcolumn}
\usepackage{bm}
\usepackage{hyperref}
\usepackage[latin9]{inputenc}
\usepackage{xspace}
\usepackage{color}

\begin{document}

\title{Circuit QED with a Nonlinear Resonator: ac-Stark Shift and Dephasing}

\author{F.~R.~Ong}
 \affiliation{Quantronics group, Service de Physique de l'État Condensé
(CNRS URA 2464), IRAMIS, DSM, CEA-Saclay, 91191 Gif-sur-Yvette, France }
 \author{M.~Boissonneault}
\affiliation{D\'epartement de Physique, Universit\'e de Sherbrooke, Sherbrooke, Qu\'ebec, Canada, J1K 2R1}
 \author{F.~Mallet}
 \affiliation{Quantronics group, Service de Physique de l'État Condensé
(CNRS URA 2464), IRAMIS, DSM, CEA-Saclay, 91191 Gif-sur-Yvette, France }
 \author{A.~Palacios-Laloy}
 \affiliation{Quantronics group, Service de Physique de l'État Condensé
(CNRS URA 2464), IRAMIS, DSM, CEA-Saclay, 91191 Gif-sur-Yvette, France }
 \author{A.~Dewes}
 \affiliation{Quantronics group, Service de Physique de l'État Condensé
(CNRS URA 2464), IRAMIS, DSM, CEA-Saclay, 91191 Gif-sur-Yvette, France }
 \author{A.~C.~Doherty}
 \affiliation{School of Mathematics and Physics, The University of Queensland, St Lucia, QLD 4072, Australia}
 \affiliation{School of Physics, The University of Sydney, Sydney, NSW 2006, Australia}
 \author{A.~Blais}
\affiliation{D\'epartement de Physique, Universit\'e de Sherbrooke, Sherbrooke, Qu\'ebec, Canada, J1K 2R1}
 \author{P.~Bertet}
 \affiliation{Quantronics group, Service de Physique de l'État Condensé
(CNRS URA 2464), IRAMIS, DSM, CEA-Saclay, 91191 Gif-sur-Yvette, France }
 \email{patrice.bertet@cea.fr}
 \author{D.~Vion}
 \affiliation{Quantronics group, Service de Physique de l'État Condensé
(CNRS URA 2464), IRAMIS, DSM, CEA-Saclay, 91191 Gif-sur-Yvette, France }
 \author{D.~Esteve}
 \affiliation{Quantronics group, Service de Physique de l'État Condensé
(CNRS URA 2464), IRAMIS, DSM, CEA-Saclay, 91191 Gif-sur-Yvette, France }

\date{\today}

\begin{abstract}
We have performed spectroscopic measurements of a superconducting qubit dispersively coupled to a nonlinear resonator driven by a pump microwave field. Measurements of the qubit frequency shift provide a sensitive probe of the intracavity field, yielding a precise characterization of the resonator nonlinearity. The qubit linewidth has a complex dependence on the pump frequency and amplitude, which is correlated with the gain of the nonlinear resonator operated as a small-signal amplifier. The corresponding dephasing rate is found to be close to the quantum limit in the low-gain limit of the amplifier.
\end{abstract}


\maketitle


In quantum mechanics, any measurement necessarily induces decoherence in the variable conjugate to the one being measured. This principle is quantitatively expressed by the inequality $\Gamma_{\phi \rm m} \geq \Gamma_{\rm meas} /2$ between the system's measurement-induced dephasing rate $\Gamma_{\phi \rm m}$ and the measurement rate $\Gamma_{\rm meas}$, stating that the most efficient detector can only measure as fast as it dephases~\cite{clerk}. Determining how far a specific detector departs from this quantum limit is a fundamental issue. The measurement of a two-level atom by a cavity to which it is non-resonantly (dispersively) coupled has been studied in detail theoretically and experimentally, first in cavity quantum electrodynamics (QED) with Rydberg atoms~\cite{brune}, then in circuit QED with superconducting qubits~\cite{blais2004,schuster,Gambetta2008}. In this situation, information about the atom state $i$ (ground state $g$ or excited state $e$) is encoded in the complex dimensionless amplitude of the qubit-dependent intracavity field $\alpha_{\rm i}$ (the {\it pointer states}~\cite{zurek2003}) and is quantified by the {\it distinguishability} $D=|\alpha_{\rm e}-\alpha_{\rm g}|$. The backaction of this dispersive readout consists in a shift of the atom frequency called the ac-Stark shift and in a gradual dephasing at a rate $\Gamma_{\phi \rm m}=\kappa D ^ 2 / 2$ ($\kappa$ is the resonator damping rate) reaching the quantum limit $\Gamma_{\rm meas}/2$~\cite{Gambetta2008}.

Introducing a nonlinearity into the measuring cavity makes it possible to turn it into an active device, which can considerably enhance the signal-to-noise ratio, a key requirement for high-fidelity measurement of superconducting qubits~\cite{vijay2010,lupascu2007,mallet}. In appropriate parameter ranges, pumped nonlinear resonators indeed behave as noiseless parametric amplifiers for small incoming signals~\cite{lehnert}, generate squeezing~\cite{castellanos-beltran2008}, and display bistability~\cite{siddiqi2004,boaknin,jbareview}. An important fundamental open question is then to determine in what respect these novel physical effects modify the backaction of the measuring cavity onto the atom, and to investigate whether the measurement is still at the quantum-limit or not.

To address this issue, we measure the spectrum of a superconducting transmon qubit~\cite{koch} dispersively coupled to a nonlinear resonator pumped by a microwave tone. Both the ac-Stark shift and the measurement-induced dephasing show new features compared to the case of a passive linear resonator. The ac-Stark shift provides a sensitive probe of the intraresonator mean photon number $\bar{n}$ and allows us to characterize the resonator nonlinearity with high precision. The measurement-induced dephasing rate is found to have a completely different dependence on the pump power than in a linear resonator: instead of growing linearly with $\bar{n}$, $\Gamma_{\phi \rm m}$ peaks where the parametric amplifier gain is maximum and decreases towards zero at large $\bar{n}$. This unexpected behavior is well described by the simple relation $\Gamma_{\phi \rm m}=\kappa D ^ 2 / 2$, $D$ being now the distinguishability between the pointer states of the {\it nonlinear} resonator. We finally discuss how close this backaction is to the quantum limit.


\begin{figure}[t]
\hspace{0mm}
\includegraphics[width=8.35cm,angle=0]{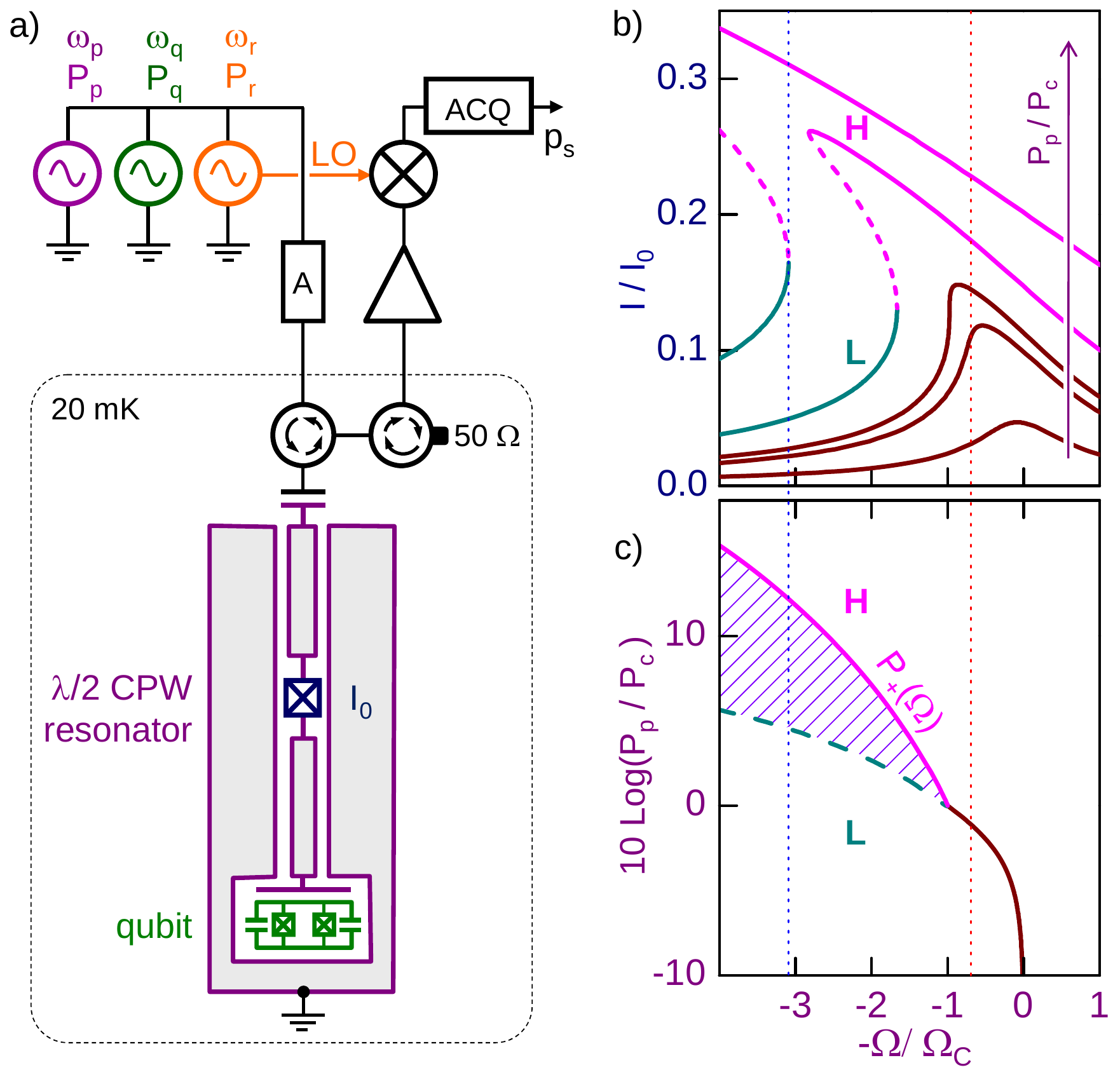}
\caption{(color online) (a)~Experimental setup: a transmon qubit is strongly coupled to a coplanar resonator made nonlinear with a Josephson junction. The sample is cooled to $20$~mK and is driven through an attenuator $A$ by three microwave sources. Source $p$ is used to establish a pump field in the resonator, source $q$ for qubit spectroscopy, and source $r$ as a JBA readout: its signal at $\omega_{\rm r}$ is reflected from the resonator and routed through circulators to a cryogenic amplifier, a demodulator, and a digitizer, which yields the JBA switching probability $p_{\rm s}$ and thus the probability of the qubit excited state  $e$. (b)~Frequency dependence of the maximum current $I$ in the resonator calculated for the sample parameters and for reduced powers $P_{\rm p}/P_{\rm c}=0.1, 0.635, 1, 3.5$, and $16.35$ (bottom to top). (c) Stability diagram of the resonator in the  $\Omega$-$P_{\rm p}$ plane. The solid lines indicate the highest PA gain below $\Omega_{\rm c}$ and the power at which the resonator bifurcates from the low- (L) to the high- (H) amplitude state, respectively. The hatched area is the bistability region. Vertical dotted lines correspond the two datasets of Figs.~2-3.}
\label{fig1}
\end{figure}

In our experiment [see Fig.~\ref{fig1}a)], a transmon qubit is capacitively coupled with a strength $g/2\pi=42.4$~MHz to a $\lambda/2$ coplanar resonator of frequency $\omega_0/2\pi=6.4535$~GHz and quality factor $Q=$~670 (damping rate $\kappa/2\pi=9.6$~MHz). The resonator is made nonlinear by inserting a Josephson junction of critical current $I_0=750$~nA in its center\cite{mallet}. We model \cite{wallquist2006,wilson2010} this nonlinear distributed resonator by one single mode with Hamiltonian $H/\hbar=\omega_0 a^\dagger a + (K/2) (a^\dagger)^2 a^2 + (K'/3) (a^\dagger)^3 a^3 $ containing a Kerr nonlinearity with constants $K$ and $K'=2 \cdot 10^{-3} K$ that can be explicitly derived from the circuit parameters (see Supplementary Material). The classical steady-state response $\alpha$ of such a Kerr resonator (KR, also called Duffing oscillator) to a pump drive of frequency $\omega_{\rm p}$ and input power $P_{\rm p}$ is therefore highly nonlinear and given by~\cite{yurke}
\begin{equation}
i \left(\Omega \frac{\kappa}{2} \alpha + K |\alpha |^2 \alpha +  K' |\alpha|^4 \alpha\right) + \frac{\kappa}{2} \alpha = -i \epsilon_{\rm p},
\label{alpha}
\end{equation}
where $\Omega=2 Q (1 - \omega_{\rm p} / \omega_0)$ is the reduced detuning of the pump and $\epsilon_{\rm p}=\sqrt{\kappa P_{\rm p} / A \hbar \omega_{\rm p}}$ its reduced amplitude ($A$ is the total attenuation of the input line). The maximum amplitude of the oscillating current $I=\sqrt{\hbar / \pi Z_0} \omega_{0} |\alpha| $ in the resonator is shown for our sample parameters in Fig.~\ref{fig1}b). At low drive amplitude, the response is a Lorentzian around $\Omega=0$ as for a linear resonator. For stronger drive, the resonance frequency shifts downwards and shows a sharpened response in a frequency window in which the resonator behaves as a parametric amplifier (PA)~\cite{lehnert}, until the slope becomes infinite for a critical power $P_{\rm c}$ at $\Omega_{\rm c}=\sqrt{3}$. For $P_{\rm p}>P_{\rm c}$ and $\Omega>\Omega_{\rm c}$, two stable solutions of different oscillation amplitude can coexist, $L$ (low) and $H$ (high). In this bistable regime called bifurcation amplification (BA), the transition from $L$ to $H$ occurs abruptly when ramping up the pump power at the bifurcation threshold $P_+(\Omega)$. Fig.~\ref{fig1}c) summarizes these properties.

The qubit-resonator detuning is fixed at $\Delta/2 \pi= 732$~MHz $\gg g/2\pi$, so that their interaction is well described by the dispersive Hamiltonian $H_{\rm int} = \hbar (\chi \sigma_{\rm z} + \bar{s} ) a^\dagger a $ with $\chi/2\pi=-0.8$~MHz and $\bar{s} / 2 \pi=1.7$~MHz~\cite{boissonneault2010}. In this dispersive regime, the resonator frequency takes the qubit-state dependent value $\omega_{\rm i}=\omega_0 + \bar{s} \pm \chi$ for $i=g,e$. Consequently, the reduced pump detuning $\Omega_{\rm i}$ as well as the bifurcation threshold $P_+(\Omega_{\rm i})$ now depend on the qubit state. This allows us to readout the qubit by sending microwave pulses with a frequency $\omega_{\rm r}/2\pi=6.439$~GHz ($\Omega_{\rm g}/\Omega_{\rm c} = 2$) and power $P_{\rm r}$ between $P_+(\Omega_{\rm e})$ and $P_+(\Omega_{\rm g})$~\cite{mallet}. After reflection on the cavity, the phase of the readout pulse is measured by homodyne detection yielding the final oscillator state. By repeating this sequence, the Josephson Bifurcation Amplifier (JBA) switching probability $p_{\rm s}$, and thus the qubit excited-state probability, is determined. 

\begin{figure}[!t]
\hspace{0mm}
\includegraphics[width=8.35cm,angle=0]{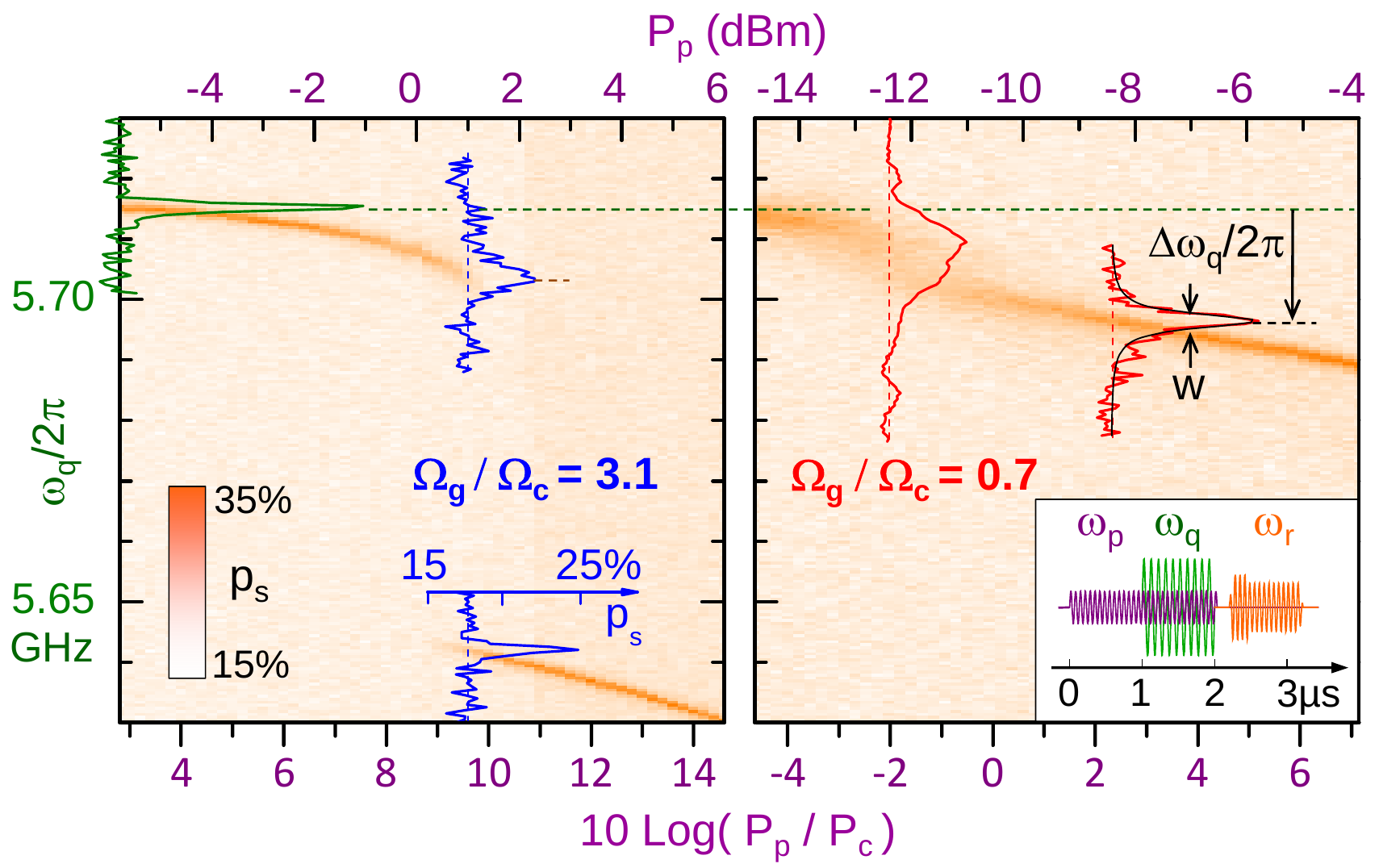}
\caption{(color online) 2D plots of the switching probability $p_{\rm s}(\omega_{\rm q},P_{\rm p})$ for $\Omega_{\rm g}/\Omega_{\rm c}=3.1$ and $0.7$ (left and right panels). A few qubit lines are shown in overlay for negligible field amplitude in the resonator (top left), near the switching point at $P_{\rm p}= 1.0$~dBm for $\Omega_{\rm g}/\Omega_{\rm c}=3.1$, and at  $P_{\rm p}=-12.4$~dBm and $-8.6$~dBm for $\Omega_{\rm g}/\Omega_{\rm c}=0.7$. The horizontal dashed line indicates the qubit frequency at zero cavity field. Lorentzian fits of the qubit lines (see example at  $-8.6$~dBm) yield the ac-stark shift  $\Delta \omega_{\rm q}$ and the FWHM linewidth $w$. Inset: microwave pulse sequence used.
}
\label{fig2}
\end{figure}


To study the backaction of the pumped KR, we perform qubit spectroscopy with a microwave pulse of varying frequency $\omega_{\rm q}$ and fixed power $P_{\rm q}$ while the cavity is driven by a microwave pump pulse of varying frequency $\omega_{\rm p}$ and power $P_{\rm p}$. The experimental sequence ends with a qubit readout pulse (see inset of Fig.~\ref{fig2}). Note that the pump pulse starts long before the spectroscopy pulse so that the intracavity field has reached its stationary state. Moreover, the readout pulse is applied $200$~ns after switching off both other pulses, a time long enough to let the intraresonator field relax before readout, but shorter than the qubit relaxation time $T_{\rm 1}=700$~ns. Fig.~\ref{fig2} shows the resulting qubit spectrum as a function of $P_{\rm p}$ at two pump frequencies above and below $\Omega_{\rm c}$. We observe the ac-Stark shift of the qubit resonance frequency towards lower values, as well as the broadening of the qubit line. This is akin to the behavior observed with a linear resonator, but with a very different dependence on $P_{\rm p}$~\cite{schuster}. First, an abrupt discontinuity in the ac-Stark shift at $\Omega_{\rm g}/\Omega_{\rm c}=3.1$ clearly indicates the sudden increase in the intracavity average photon number $\bar{n}$ as the resonator switches from $L$ to $H$. In this region, two spectroscopic peaks are observed at a given pump power. Second, the linewidth {\it narrows down} at large $\bar{n}$, in strong contrast with the linear resonator case where it increases linearly~\cite{schuster}.  

\begin{figure}[!h]
\begin{center}
\hspace{0mm}
\includegraphics[width=8.35cm,angle=0]{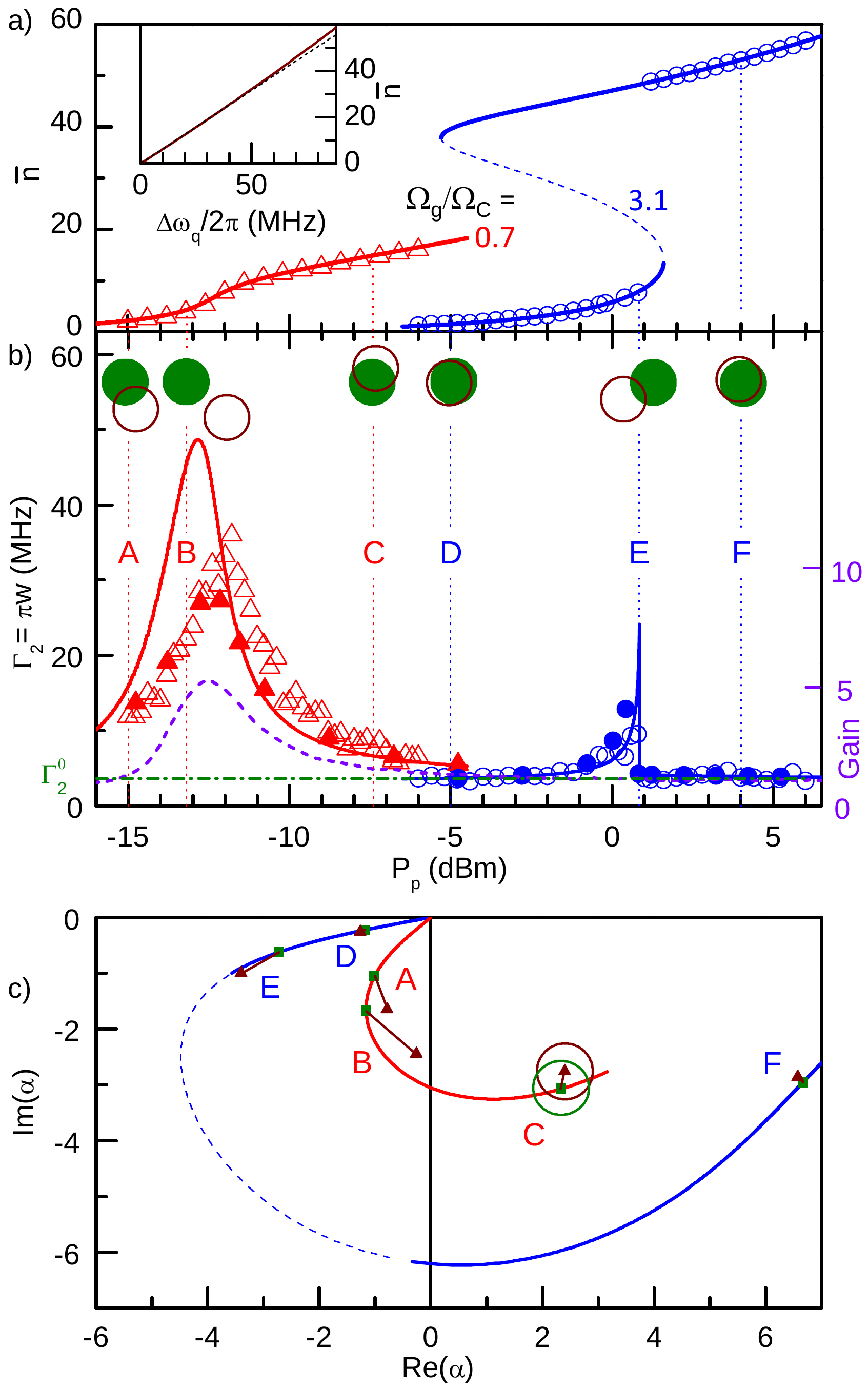}
\caption{(color online) (a) Experimental (open symbols) average photon number $\bar{n}$ and corresponding fits (lines - see text) as a function of $P_{\rm p}$ for the same datasets  $\Omega_{\rm g}/\Omega_{\rm c}=3.1$ (circles) and $\Omega_{\rm g}/\Omega_{\rm c}=0.7$ (triangles) as in  Fig.~\ref{fig2}. Experimental points are obtained by converting the measured $\Delta \omega_{\rm q}$ according to the ac-Stark shift model of the inset (solid line - see text ; the dashed line shows the linear approximation). (b)~Qubit dephasing rate $\Gamma_2$  measured for the same datasets (open symbols) and calculated either by numerical integration of the system master equation (solid symbols) or using Eq.~\eqref{gammam} (solid lines). The horizontal dashed-dotted line indicates the intrinsic dephasing at zero field. The measured parametric power gain  (dashed line) is also shown  for comparison for $\Omega_{\rm g}/\Omega_{\rm c} = 0.7$. (c)~Complex cavity field amplitude $\alpha_{\rm g}$ (solid and dashed lines) calculated from Eq.~\eqref{alpha} for increasing pump powers and for the same datasets. Both $\alpha_{\rm g}$ (squares) and $\alpha_{\rm e}$ (triangles) are also shown at six points labelled A-F. Their separation (segments) is to be compared to the uncertainty disk of a coherent state (open circles or disks shown at point C and at points A-F in panel b).}
\label{fig3}
\end{center}
\end{figure}

Our measurements of the qubit frequency shift $\Delta \omega_{\rm q} $ versus $P_{\rm p}$ allow us to quantitatively determine $\bar{n}$, using a model for the ac-Stark shift that includes the next order correction to the linear ac-Stark shift formula $\bar{n}=\Delta \omega_{\rm q}/(2\chi)$~\cite{boissonneault2008,*boissonneault2009} and that takes into account the first 5 levels of the transmon ~\cite{koch,boissonneault2010} (see Fig.~\ref{fig3}a). We then fit the resulting experimental $\bar{n}(P_{\rm p})$ curves with the $\bar{n}$ values calculated from the sole dynamics of the resonator (with the qubit in $g$) given by the square modulus of the solutions of Eq.~\eqref{alpha} using $K$ and $A$ as the only fitting parameters. The agreement is excellent over the whole $(\Omega,P_{\rm p})$ range for $K/2\pi=- 625\pm 15$~kHz and $A= 110.8\pm 0.2$~dB, which is consistent with the design value $K/2\pi = -750 \pm 250$~kHz and with independent measurements of the line attenuation $A=111\pm 2$~dB. This demonstrates that measuring the ac-Stark shift of a qubit is a sensitive method for probing the field inside a nonlinear resonator and for characterizing its Kerr nonlinearity~\cite{lehnert}.

\noindent


We show in Fig.~\ref{fig3}b) the qubit total dephasing rate $\Gamma_2 = \pi w $, with $w$ the full width at half maximum (FWHM) of a Lorentzian fit to the experimental data. In addition to the field-induced dephasing $\Gamma_{\phi \rm m}$, $\Gamma_2$ also includes a constant contribution $\Gamma_2^0$ due to all other dephasing processes (mainly energy relaxation, flux noise and radiative broadening). For the sake of comparison, we also show the independently measured gain $G$ of the KR operated as a parametric amplifier on a small additional signal frequency-shifted by $100$~kHz from $\omega_{\rm p}/2\pi$. Below and above $\Omega_{\rm c}$, $\Gamma_2 (P_{\rm p})$ peaks where $G$ is maximum and near the bifurcation threshold respectively, and tends towards $\Gamma_2^0$ at large $\bar{n}$.

To account for this behavior, we have calculated $\Gamma_2$ by numerical integration of the master equation of the multi-level transmon coupled to the KR with a Jaynes-Cummings Hamiltonian for the sample parameters and damping rates determined independently [see full symbols in Fig.~\ref{fig3}b)]. The agreement with the experimental data is excellent with no adjustable parameter, in contrast with experiments with a flux-qubit in which an unexplained shortening of the qubit relaxation time is observed when the KR is driven above bifurcation~\cite{picot,serban}. We also derive an analytical expression for $\Gamma_{\phi \rm m}$ starting from the master equation in the dispersive limit, using methods similar to those of \cite{boissonneault2009} for the linear case. This is done by linearizing the quantum fluctuations of the field around the two classical solutions $\alpha_{\rm i}$ of Eq.~\eqref{alpha} that correspond to the two qubit states, assuming that the separation between these pointer states is small and neglecting squeezing and transients ($\Gamma_{\phi \rm m} \ll \kappa$). The details of these calculations, which go beyond a simple linear response approximation, will be presented elsewhere~\cite{maxime}. In this way, we obtain

\begin{equation}
\Gamma_{\phi \rm m}=\frac{\kappa}{2} |\alpha_{\rm e}-\alpha_{\rm g}|^2=\frac{\kappa}{2} D^2.
\label{gammam}
\end{equation}

\noindent
We thus recover the same link as for a linear resonator between decoherence rate and distinguishability. Here however, due to the nonlinearity, $D$ and therefore $\Gamma_{\phi \rm m}$ are no longer proportional to $\bar{n}$~\cite{schuster} and on the contrary tend towards zero at large $\bar{n}$. Since in a pumped KR the susceptibility $| \partial \alpha / \partial \Omega |$ varies similarly to the small-signal gain $G$ \cite{manucharyan,clerk2010}, Eq.~\eqref{gammam} also explains the correlation observed between $\Gamma_{\phi \rm m}$ and $G$. The prediction of Eq.~\eqref{gammam} calculated without any adjustable parameter as well as the corresponding pointer states are shown in Fig.~\ref{fig3}b) and c). The theory reproduces well the non-trivial dependence of $\Gamma_2$ on the pump power over a large parameter range, and in particular the dephasing peak observed above and below $\Omega_{\rm c}$. The direct correlation between $\Gamma_2$ and $D$ is clear. The agreement is quantitative in the validity range of our approximation $D\ll 1$, which is satisfied by our data when $\Gamma_2$ is below $\sim 10$~MHz.

In order to verify whether the pumped KR backaction is at the quantum limit, $\Gamma_{\phi \rm m}$ needs to be compared to the rate $\Gamma_{\rm meas}$ at which information about the atom state leaks out of the resonator. If the two pointer states are coherent states, $\Gamma_{\rm meas}=\kappa D^2$ \cite{Gambetta2008} and our results therefore establish that the quantum limit is reached. According to the quantum theory of the pumped KR \cite{drummondwalls,clerk2010}, this is however true only in the limit of small parametric gain $G \sim 1$ [corresponding to $P_p$ significantly different from $P_+(\Omega)$]. Indeed, when $G \gg 1$, the intracavity field is expected to show enhanced phase-dependent fluctuations and the field reflected on the cavity to show some degree of squeezing. The measurement rate then needs to be re-evaluated. Note that recent theoretical work \cite{clerk2010} showed that if the linear response theory is valid and if the gain is large, the backaction of the pumped KR misses the quantum limit by a large factor of order $G$. In contrast our experimental parameters, which are typical for circuit QED, require going beyond linear response but involve limited parametric gain. It is also important to note that our work does not address the situation encountered in the BA regime close to the bifurcation threshold $P_+(\Omega)$, in which $\alpha_{\rm g,e}$ are the two different oscillator states $L,H$. As a result, further work is needed to decide whether the readout of a qubit by a JBA is quantum-limited or not.


In summary, we have measured the quantum backaction of a pumped nonlinear resonator on a dispersively coupled qubit. We observe a non-trivial dependence of the measurement-induced dephasing on the pump power, which we link to the distinguishability between the pointer states of the nonlinear resonator and to the gain of the resonator operated as a parametric amplifier. In the small gain limit, we find that the backaction is quantum-limited; further theoretical work is needed to describe the high gain regions. More generally, our results demonstrate that circuit QED is an ideal playground to study the interplay between strong coupling and nonlinear effects such as bistability, squeezing or parametric amplification.

\begin{acknowledgments}
We acknowledge discussions with M.~Dykman, M.~Devoret, J.~Gambetta, R.~Vijay and within the Quantronics group as well as technical support from P.~S\'enat, P.-F.~Orfila, T.~David, and J.-C.~Tack. We acknowledge support from NSERC, the Alfred P. Sloan Foundation, CIFAR, the ANR project Quantjo, the European project SCOPE and the Australian Research Council.
\end{acknowledgments}


\end{document}


\title{Supplementary Material for ``Circuit QED with a Nonlinear Resonator : ac-Stark Shift and Dephasing"'}

\author{F.~R.~Ong}
 \affiliation{CEA-Saclay, Gif-sur-Yvette, France}
 \author{M.~Boissonneault}
\affiliation{D\'epartement de Physique, Universit\'e de Sherbrooke, Sherbrooke, Qu\'ebec, Canada, J1K 2R1}
 \author{F.~Mallet}
 \affiliation{CEA-Saclay, Gif-sur-Yvette, France}
 \author{A.~Palacios-Laloy}
 \affiliation{CEA-Saclay, Gif-sur-Yvette, France}
 \author{A.~Dewes}
 \affiliation{CEA-Saclay, Gif-sur-Yvette, France}
 \author{A.~C.~Doherty}
 \affiliation{School of Mathematics and Physics, The University of Queensland, St Lucia, QLD 4072, Australia}
 \affiliation{School of Physics, The University of Sydney, Sydney, NSW 2006, Australia}
 \author{A.~Blais}
\affiliation{D\'epartement de Physique, Universit\'e de Sherbrooke, Sherbrooke, Qu\'ebec, Canada, J1K 2R1}
 \author{P.~Bertet}
 \affiliation{CEA-Saclay, Gif-sur-Yvette, France}
 \author{D.~Vion}
 \affiliation{CEA-Saclay, Gif-sur-Yvette, France}
 \author{D.~Esteve}
 \affiliation{CEA-Saclay, Gif-sur-Yvette, France}

\date{\today}

\maketitle

In this Supplementary Material, we obtain analytical expressions for the Kerr constants appearing in the Hamiltonian of a nonlinear resonator. It consists of a $\lambda/2$ coplanar waveguide resonator with an embedded Josephson junction of critical current $I_0$ embedded in the middle conductor. We first discuss why it is appropriate to model the distributed nonlinear resonator by a single mode with Kerr nonlinearities. Indeed, in doing so we neglect the higher frequency modes of the resonator. This is justified in our case because all the resonator pumping is performed very near the Kerr resonator frequency $\omega_0$, and because the resonance frequency of the other modes of our resonator is far from being a multiple of $\omega_0$ as discussed in \cite{wallquist2006}, allowing us to neglect cross-Kerr coupling between various modes \cite{wilson2010}.

\begin{figure}[t]
\includegraphics[width=8.35cm,angle=0]{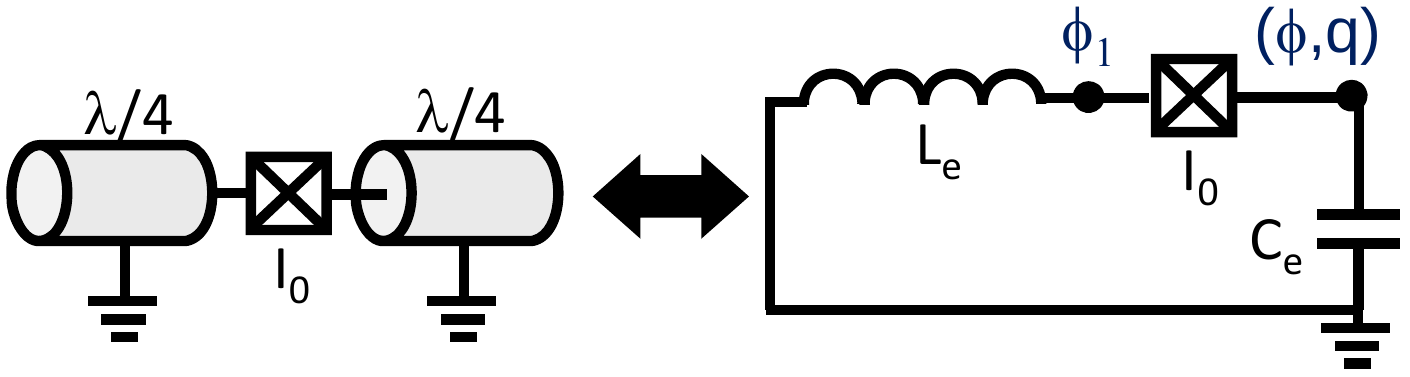}
\caption{Equivalence between the distributed nonlinear resonator and a series combination of an equivalent inductance $L_{\rm e}$, capacitance $C_{\rm e}$ and Josephson junction of critical current $I_0$.
}
\label{figSM}
\end{figure}

To derive the Kerr constants, the first step is to map the distributed nonlinear resonator onto an equivalent series combination of a lumped element inductance $L_{\rm e}$, capacitance $C_{\rm e}$ (which includes the junction capacitance) and Josephson junction of critical current $I_0$, as shown in Fig.~\ref{figSM}. Note that this mapping is only possible here because we can neglect cross-Kerr terms between the various modes of the resonator sin
In the limit where the Josephson inductance $L_{\rm J}$ is completely negligible compared to the resonator inductance, one can show that $L_{\rm e} = \pi Z_0 / 2 \omega_1$ and $C_{\rm e} = 2 / \pi Z_0 \omega_1$, where $Z_0 \approx$~50 $\Omega$ is the resonator characteristic impedance and $\omega_1$ the resonance frequency in absence of the junction. When $L_{\rm J}$ is not negligible, one has to adjust numerically  $L_{\rm e}$ and $C_{\rm e}$ for the impedance of the equivalent circuit to fit the exact distributed resonator impedance. This is the approach that has been used in the Letter.
 
Using the notation introduced in Fig.~1, we obtain the equivalent circuit Hamiltonian
\begin{equation}
H=\frac{\phi_{1}^{2}}{2L_{\rm e}}-E_{\rm J}\cos\left(\frac{\phi-\phi_{1}}{\varphi_{0}}\right)+\frac{q^{2}}{2C_{\rm e}}.\end{equation}
Since the current $I$ flowing through the inductance and the junction is the same, we also have 
\begin{equation}
I=\frac{\phi_{1}}{L_{\rm e}}=I_{0}\sin\left(\frac{\phi-\phi_{1}}{\varphi_{0}}\right),
\end{equation}
yielding an implicit relation $\phi_{1}=g(\phi)$ between the two phases. Eliminating $\phi_1$, the Hamiltonian thus takes the form
\begin{equation}
H=\frac{\left[g(\phi)\right]^{2}}{2L_{\rm e}}-E_{\rm J}\cos\frac{\phi-g(\phi)}{\varphi_{0}}+\frac{q^{2}}{2C_{\rm e}}.
\end{equation}

By expanding $g(\phi)$ in powers of $\phi$, we obtain the nonlinear resonator Hamiltonian to any order of Josephson junction nonlinearity. For instance, to fourth order we find
\begin{equation}
H=\frac{\phi^{2}}{2L_{\rm t}}+\frac{q^{2}}{2C_{\rm e}}-\frac{1}{24}p^{3}\frac{\phi^{4}}{L_{\rm t}\varphi_{0}^{2}}\end{equation}
where $L_{\rm t}=L_{\rm J} + L_{\rm e}$ is the total inductance and $p=L_{\rm J} / L_{\rm t}$. 

This Hamiltonian can be writen in terms of the creation $a^\dag$ and annihilation $a$ operators with $\phi=i \sqrt{\hbar Z_{\rm e}/2} (a-a^{\dagger})$ and $q =\sqrt{\hbar /2 Z_{\rm e}} (a+a^{\dagger})$ with  $Z_{\rm e}=\sqrt{L_{\rm t} / C_{\rm e}}$.  Once expanded, the nonlinear term $\phi^{4}$ yields products of creation and annihilation operators to various powers. Using the rotating-wave approximation, we keep only those with equal number of annihilation and creation operators yielding

\begin{equation}
H=\hbar\omega_0a^{\dagger}a+\hbar\frac{K}{2}(a^{\dagger})^{2}a^{2}\label{eq:CBA Hamiltonian with K},
\end{equation}
where
\begin{equation}
K=-\frac{\pi p^{3}\omega_0Z_{\rm e}}{R_{\rm K}}
\end{equation}
is the Kerr constant, $\omega_0 = 1/\sqrt{L_{\rm t} C_{\rm e}}$ and $R_K = h/e^2$. Here and below, we have dropped the small correction to $\omega_0$ coming from $K$.

In the same way, the Hamiltonian can be expanded
to the next order of junction nonlinearity yielding the next-order Kerr constant $K'$
\begin{equation}
H=\hbar\omega_0a^{\dagger}a+\hbar\frac{K}{2}(a^{\dagger})^{2}a^{2}+\hbar\frac{K'}{3}(a^{\dagger})^{3}a^{3},
\label{eq:CBA Hamiltonian with K and K'}
\end{equation}
with
\begin{equation}
K' = \frac{2}{3p}\frac{K^2}{\omega_0}\left(10p-9\right).
\end{equation}
For our sample's parameters, this yields an approximate relation $K' \approx 0.002 K$ that we consider fixed throughout the Letter and in particular in the fit of the ac-Stark shift data.
